\documentclass[12pt,preprint]{aastex}

\shorttitle{Induced Radio and X-ray Emission}
\shortauthors{Prigara}

\begin{document}

\title{Induced Radio and X-ray Emission From an Accretion Disk}
\author{F.V.Prigara}
\affil{Institute of Microelectronics and Informatics, Russian
Academy of Sciences, 21 Universitetskaya, Yaroslavl 150007,
Russia}
\email{fprigara@imras.yar.ru}

\begin{abstract}
Radio and X-ray emission from compact sources with accretion disks
(active galactic nuclei, pulsars and X-ray binaries) is
considered. It is shown that both radio and X-ray emission from
these sources can be interpreted as emission of a hot ($T
\geqslant 3 \times 10^{7}K$) plasma in the inner part of an
accretion disk or in the disk corona. Radio emission is produced
by the maser amplification of thermal radio emission in a hot
plasma and corresponds to the transitions between highly located
energy levels (somewhat similarly to the recombination lines). The
X-ray emission is either thermal radiation from dense filaments or
is produced by the coherent inverse Compton scattering of radio
photons in the same dense filaments. The same mechanism, which
gives rise to the maser amplification at radio wavelengths,
produces also laser amplification in optical in X-ray binaries.
\end{abstract}


\section{Introduction}

In the unified model of compact radio sources, radio emission from
active galactic nuclei (AGNs) and pulsars is treated as thermal
radiation from an accretion disk amplified by a maser mechanism
\citep{pr03a}. A maser amplification of thermal radio emission in
continuum produces the high brightness temperatures of compact
radio sources and a rapid variability of total and polarized flux
density, that is characteristic for non-saturated maser sources.
In particular, pulsars signals show a variability on every
observable time scale up to nanoseconds \citep{edw03,viv01}. The
brightness temperatures of OH masers have the magnitude $T_{b}
\leqslant 10^{12}K$, and those of water masers have the magnitude
$T_{b} \leqslant 10^{15}K$ \citep{boc92}. Compact extragalactic
sources (AGNs) exhibit brightness temperatures in the range of
10$^{10}$ K to 10$^{12}$ K \citep{bow98,kel98}, so these
temperatures have an order of magnitude of those of OH masers.

Maser amplification in continuum is closely connected with the stimulated
origin of thermal radio emission. The induced origin of thermal radio
emission follows from the relations between Einstein's coefficients for a
spontaneous and induced emission of radiation. However, the detailed
mechanism of maser amplification has been unknown so far. In this paper we
show that maser amplification is produced by the inversion of the high
energy level population in a hot plasma.

The unified model of compact sources can be extended to account
for emission in other bands. X-ray binaries have roughly
two-component X-ray spectra with a thermal blackbody component and
a power law spectrum \citep{fal03}. The power law spectrum in the
X-ray range has been also detected in some radio pulsars, X-ray
pulsars and AGNs \citep{cha01}. The unified model predicts photon
indices of the power law spectrum in the X-ray range which may be
compared with the observed indices.

\section{The gaseous disk model}

It was shown recently \citep{pr03b} that thermal radio emission
has a stimulated character. According to this conception thermal
radio emission from non-uniform gas is produced by an ensemble of
individual emitters. Each of these emitters is an elementary
resonator the size of which has an order of magnitude of mean free
path \textit{l} of photons

\begin{equation}
\label{eq1}
l = \frac{{1}}{{n\sigma} }
\end{equation}

\noindent
where \textit{n} is the number density of particles and $\sigma $ is the
absorption cross-section.

The emission of each elementary resonator is coherent, with the
wavelength

\begin{equation}
\label{eq2}
\lambda = l,
\end{equation}

\noindent
and thermal radio emission of gaseous layer is incoherent sum of radiation
produced by individual emitters.

The condition (\ref{eq2}) implies that the radiation with the wavelength $\lambda $
is produced by the gaseous layer with the definite number density of
particles \textit{n} .

The condition (\ref{eq2}) is consistent with the experimental
results by Looney and Brown on the excitation of plasma waves by
electron beam \citep{che84,ale03}. The wavelength of standing wave
with the Langmuir frequency of oscillations depends on the density
as predicted by equation (\ref{eq1}). The discrete spectrum of
oscillations is produced by the non-uniformity of plasma and the
readjustment of the wavelength to the length of resonator. From
the results of experiment by Looney and Brown the absorption
cross-section for plasma can be evaluated.

The product of the wavelength by density is weakly increasing with the
increase of density. This may imply the weak dependence of the size of
elementary resonator in terms of the wavelength upon the density or,
equivalently, wavelength.

In the gaseous disk model, describing radio emitting gas nebulae
\citep{pr03a}, the number density of particles decreases
reciprocally with respect to the distance \textit{r} from the
energy center

\begin{equation}
\label{eq3}
n \propto r^{ - 1}.
\end{equation}

Together with the condition for emission (\ref{eq2}) the last equation leads to the
wavelength dependence of radio source size:

\begin{equation}
\label{eq4}
r_{\lambda}  \propto \lambda .
\end{equation}

The relation (\ref{eq4}) is indeed observed for sufficiently
extended radio sources. For example, the size of radio core of
galaxy M31 is 3.5 arcmin at the frequency 408 MHz and 1 arcmin at
the frequency 1407 MHz \citep{sha82}.

\section{Radio emission from the gaseous disk}

The spectral density of flux from an extended radio source is given by the
formula

\begin{equation}
\label{eq5}
F_{\nu}  = \frac{{1}}{{a^{2}}}\int\limits_{0}^{r_{\lambda} }  {B_{\nu} }
\left( {T} \right) \times 2\pi rdr \quad ,
\end{equation}

\noindent
where \textit{a} is a distance from radio source to the detector of
radiation, and the function $B_{\nu}  \left( {T} \right)$ is given by the
Rayleigh-Jeans formula

\begin{equation}
\label{eq6}
B_{\nu}  = 2kT\nu ^{2}/c^{2},
\end{equation}

\noindent
where $\nu $ is the frequency of radiation, \textit{k} is the Boltzmann
constant, and \textit{T} is the temperature.

. The extended radio sources may be divided in two classes. Type 1 radio
sources are characterized by a stationary convection in the gaseous disk
with an approximately uniform distribution of the temperature \textit{T$
\approx $const} giving the spectrum

\begin{equation}
\label{eq7}
F_{\nu}  \approx const \quad .
\end{equation}

Type 2 radio sources are characterized by outflows of gas with an
approximately uniform distribution of gas pressure \textit{P=nkT$ \approx
$const}. In this case the equation (\ref{eq3}) gives

\begin{equation}
\label{eq8}
T \propto r,
\end{equation}

\noindent
so the radio spectrum, according to the equation (\ref{eq5}), has the form

\begin{equation}
\label{eq9}
F_{\nu}
 \propto \nu ^{ - 1}.
\end{equation}

Both classes include numerous galactic and extragalactic objects.
In particular, edge-brightened supernova remnants \citep{kul93}
belong to the type 2 radio sources in accordance with the relation
(8), whereas center-brightened supernova remnants belong to the
type 1 radio sources.

The relationship between linear size and turnover frequency in
type 2 radio sources (gigahertz-peaked spectrum sources and
steep-spectrum sources) \citep{nag02} is a consequence of the
wavelength dependence of radio source size. The turnover frequency
is determined by the equation $r_{\nu} = R$, where R is the radius
of a gaseous disk. The same equation determines a turnover
frequency for planetary nebulae \citep{pr03a,pot84,sio01}.

\section{The unified model of compact radio sources}

The unified model of compact radio sources \citep{pr03a} invokes
an accretion disk with convection and outflows. The density
profile $n \propto r^{ - 1/2}$used in the unified model is
standard for an outflow or the convection dominated accretion flow
(CDAF) models \citep{nag01}. Here \textit{r} is the distance from
the central energy source. The temperature profile is virial, i.e.
$T \propto r^{ - 1}$, only in active galactic nuclei (AGNs). In
pulsars it has other forms. The fixed ratio of magnetic to gas
pressure is inferred for AGNs, in pulsars the decouplement of
magnetic and gas pressure is observed.

The unified model suggests the uniform maser amplification of thermal radio
emission in continuum. In the model, the inferred brightness temperatures of
radio pulsars are comparable to those of water vapor, OH and SiO masers. The
unified model predicts the flat radio spectrum for a core emission from
AGNs, the spectrum (\ref{eq9}) for an outflow, and the intermediate ($0 < \alpha <
1$) values of the spectral indices for unresolved sources, depending on the
relative contributions of an outflow and accretion disk. Here $F_{\nu}
\propto \nu ^{ - \alpha} $is the flux density and í is the frequency.

The model includes also radiation-induced solitary waves in an
accretion disk to explain the phenomena seen in pulsars
\citep{pr04}.

In the case of compact radio sources instead of the relationship (\ref{eq4}) the
relationship

\begin{equation}
\label{eq10}
r_{\lambda}  \propto \lambda ^{2}
\end{equation}

\noindent is observed \citep{lo93,lo82}. This relationship may be
explained by the effect of a gravitational field on the motion of
gas which changes the equation (\ref{eq3}) for the equation
\begin{equation}
\label{eq11}
 n \propto r^{-1/2}
\end{equation}
The mass conservation in an outflow or inflow of gas gives
\textit{nvr=const,} where \textit{v} is the velocity of flow. In
the gravitational field of a central energy source the energy
conservation gives

\begin{equation}
\label{eq12} v = \left( {v_{0}^{2} + c^{2}r_{s} /r} \right)^{1/2}
\end{equation}

\noindent
where $r_{s} $ is the Schwarzschild radius. Therefore, at small values of
the radius the equation (\ref{eq6}) is valid, whereas at the larger radii we obtain
the equation (\ref{eq3}).

It is well known that the delay of radio pulses from pulsars at
low frequencies is proportional to $\lambda ^{2}$. This fact is a
mere consequence of Eq.(\ref{eq10}), if we only assume the
existence of the radial density wave travelling across the radius
with a constant velocity and triggering the pulse radio emission.
In this treatment the pulsars also obey the $\lambda ^{2}$
dependence of compact source size. Note that the wavelength
dependence of a pulse duration is a similar effect.

The spatial distribution of SiO, water, and OH masers (each of
which emits in its own wavelength) in the maser complexes also is
consistent with the $\lambda ^{2}$ dependence of compact source
size \citep{boc92,eis02}.

To summarize, extended radio sources are characterized by the relation (\ref{eq4}),
and compact radio sources obey the relation (\ref{eq10}).

\section{Maser amplification in compact radio sources}

Recently, the energy distribution of atoms in the field of thermal
black body radiation was obtained \citep{pr03b} in the form

\begin{equation}
\label{eq13} N/N_{0} = \sigma _{a} \omega ^{2}/\left( {2\pi c^{2}}
\right)\left( {exp\left( {\hbar \omega /kT} \right) - 1} \right),
\end{equation}

\noindent
where $N_{0} $ is the population of the ground state $E_{0} $, \textit{N} is
the population of the energy level $E = E_{0} + \hbar \omega $, $\sigma _{a}
$ is the absorption cross-section, $\hbar $ is the Planck constant, and
\textit{T} is the radiation temperature.

This distribution is valid in the range $\hbar \omega /kT
\geqslant 1$, since in the limit $\hbar \omega /kT \to 0$ the line
width is going to infinity, that indicates the violation of the
one-particle approximation used by \citet{pr03b}.

The function (\ref{eq13}) has a maximum at $\hbar \omega _{m} =
1.6kT$. When the temperature exceeds the critical value of $T_{0}
= 3 \times 10^{7^{}}K$ (the inversion temperature), the population
of the energy level \textit{E} exceeds the population of the
ground state $E_{0} $. Since the function (\ref{eq13}) is
increasing in the range $\omega < \omega _{m} $ , the inversion of
the energy level population is produced also in some vicinity of
$\omega _{m} $ (below $\omega _{m} $). This suggests the maser
amplification of thermal radio emission in continuum by a hot
plasma with the temperature exceeding the critical value $T_{0} $.
Maser amplification in compact radio sources was assumed earlier
by \citet{pr03a} based on the high brightness temperatures of
AGNs. Since a hot plasma in an accretion disk is concentrated
nearby the central energy source, maser amplification is
characteristic for compact radio sources.

It is clear that, when the temperature of a plasma is below $T_{0}
$, the radio flux is very small, and when the temperature exceeds
$T_{0} $, radio emission is on. This an on-off cycle is detected
in the radio pulsar PSR B1259-63 \citep{qia03}. Similar is an
on-off cycle in X-ray pulsars, e.g., the 35-day cycle in Her X-1.
It implies that X-ray emission from X-ray pulsars is produced by
the laser amplification in continuum which is quite analogues to
maser amplification at radio wavelengths.

\section{The photon indices of X-ray emission}

X-ray binaries normally have the two most pronounced states
\citep{fal03}. The first one is the high/soft state dominated by
the thermal blackbody emission from a thin disk. The second one is
the low/hard state which is characterized by a dominant hard
power-law spectrum whereas the thermal spectrum is weak or absent.
The power-law spectrum is commonly attributed to an optically thin
accretion flow or disk corona. However, there are other plausible
origins of the power-law spectrum \citep{fal03}.

The power-law spectrum in the X-ray range has also been detected
in young classical pulsars, anomalous X-ray pulsars (AXPs) and
soft gamma-ray repeaters (SGRs) \citep{cha01}. All these objects
are believed to be isolated neutron stars. For AXPs and SGRs, the
fallback accretion disk model has been proposed \citep{qia03}. It
is assumed that the accretion is fed by fallback material after
the original supernova explosion.

To characterize the power-law spectrum, we introduce the photon index,
$\Gamma $, defined such that photon number flux $dN/dE \propto E^{ - \Gamma
}$. Since $E = h\nu $, where \textit{h} is the Planck constant, $\Gamma - 1$
is the spectral index in the X-ray range.

Classical young pulsars have power-law X-ray spectra with $\Gamma = 1.1 -
1.7$ in the 0.1-10 keV band, that corresponds to the spectral indices
$\Gamma - 1 = 0.1 - 0.7$. Thus, the spectral indices for these pulsars lie
between 0 and 1, similar to the spectral indices of radio emission from
supernova remnants (SNRs).

The X-ray spectrum of most AXPs is best characterized by a
two-component spectrum consisting of a $\sim $0.5 keV blackbody
emission and a steep ($\Gamma = 3 - 4$) power law spectrum, with
comparable luminosities in both components \citep{cha01}. The
spectral indices for AXPs are in the same range, $2 < \Gamma - 1 <
3$, as the spectral indices of radio emission from pulsars . The
latter can be inferred by making use of the density profile of
compact radio sources, similarly to the spectra (\ref{eq7}) and
(\ref{eq8}).

Active galactic nuclei (AGNs) typically lie within a range of photon indices
$\Gamma = 1.2 - 2.2$, so the spectral indices $\Gamma - 1 = 0.2 - 1.2$ are
close to those of classical young pulsars, i.e. lie in the same range as the
spectral indices of radio emission from extended sources.

The theory of thermal emission from a gas with account for stimulated
radiation processes gives the two possibilities to explain these values of
photon indices in the X-ray range. The first one is to apply the above
theory to the X-ray emission from a hot, optically thin in a classical sense
disk corona. If the temperature of a gas is sufficiently high, then the
Rayleigh-Jeans formula (\ref{eq6}) is still valid, and the only difference with the
radio band is the order of magnitude of the density, \textit{n}, required by
the equations (\ref{eq1}) and (\ref{eq2}) to produce X-ray emission. In this scheme, the
power law spectrum can be produced either by the hot inner disk corona
($\Gamma = 3 - 4$) or by the hot filaments in the thick outer disk with
outflows ($\Gamma = 1 - 2$).

Another possibility is the inverse Compton scattering of radio
photons from an accretion disk by the hot disk corona or the hot
filaments. However, the incoherent Compton scattering is not
relevant in this case, because it does not conserves the spectral
indices. In fact, the final spectrum is determined mostly by the
spectral energy distribution of electrons and weakly depends upon
the original radio spectrum. We should assume, instead, the
coherent Compton scattering to reproduce the original spectral
indices of radio emission (cf. \citep{ree82}). It is plausible,
that both these processes, the emission of thermal radiation and
the inverse Compton scattering, contribute to the observed
spectra.

The coherent inverse Compton scattering of radio photons from an
accretion disk is supported by a strong (one-to-one) correlation
between radio oscillation events and series of spectrally hard
states in GRS 1915+105 \citep{kle02}. Another observational
evidence for the inverse Compton scattering in the hot disk corona
is that the spectral indices of radio emission from X-ray binaries
correspond to the emission from an outer disk. The radio emission
from the inner part of an accretion disk, which has been detected
in pulsars, in X-ray binaries is absent. It suggests that this
emission is converted into X-ray emission via inverse Compton
scattering.

\section{Conclusions}

In this paper, we elucidate the mechanism of maser amplification
in compact radio sources, which has been suggested earlier based
on the high brightness temperatures of these sources. Maser
amplification is produced by the inversion of the high energy
level population in a hot plasma.The inversion of the level
population can produce also laser amplification in optical (e.g.,
the high variable shifts and intensities of the weak emission
lines in Sco X-1 can be attributed to the weak laser sources) and
X-ray bands. An on-off cycle in radio and X-ray pulsars may be
explained by the periodic changes of the emitting gas temperature
from higher to lower than the inversion temperature values.

The photon indices of the power law spectra in the X-ray range are
obtained similar to the spectral indices of radio emission in the
unified model of compact radio sources. The only difference is the
higher density of an emitting gas. However, the detected
correlation between radio and X-ray emission in X-ray sources
suggests that another mechanism (the coherent inverse Compton
scattering of radio photons) also contributes to the observed
spectra.

\textit{Acknowledgements.} The author is grateful to
D.A.Kompaneets, Y.Y.Kovalev, V.N.Lukash, B.E.Stern, and N.A.Tsvyk
for useful discussions.

\end{document}